\documentclass[11pt]{article}
\usepackage{hyperref}
\pdfoutput=1
\usepackage{amsmath,amssymb}              
\usepackage{authblk}                      
\usepackage{graphics,graphicx}            
\usepackage{subfigure}                    
\usepackage{import}                       
\usepackage{color}                        
\usepackage{booktabs}                     
\newcommand{\mm}{\ensuremath{\mathrm{mm}}}

\newcommand{\ml}{\ensuremath{\mathrm{ml}}}

\newcommand{\SU}{\ensuremath{\mathrm{Su^+}}}		
\newcommand{\WE}{\ensuremath{\mathrm{We^+}}}
\newcommand{\g}{\ensuremath{g_\mathrm{o}}}

\begin{document}
\title{Spontaneous Capillarity-Driven Droplet Ejection}
\author[1]{Andrew Wollman}
\affil[1]{Portland State University}
\author[2]{Trevor Snyder}
\affil[2]{Xerox Inc.}
\author[3]{Donald Pettit}
\affil[3]{NASA, Johnson Space Center}
\author[1]{Mark Weislogel}


\maketitle

\begin{abstract}
The first large length-scale capillary rise experiments were conducted by R. Siegel using a drop tower at NASA LeRC shortly after the 1957 launch of Sputnik I.  Siegel was curious if the wetting fluid would expel from the end of short capillary tubes in a low-gravity environment. He observed that although the fluid partially left the tubes, it was always pulled back by surface tension, which caused the fluid to remain pinned to the tubes' end. By exploiting tube geometry and fluid properties, we demonstrate that such capillary flows can in fact eject a variety of jets and drops. This fluid dynamics video provides a historical overview of such spontaneous capillarity-driven droplet ejection. Footage of terrestrial and low earth orbit experiments are also shown. Droplets generated in a microgravity environment are $10^6$ times larger than those ejected in a terrestrial environment. The accompanying article provides a summary of the critical parameters and experimental procedures. Scaling the governing equations reveals the dimensionless groups that identify topological regimes of droplet behavior which provides a novel perspective from which to further investigate jets, droplets, and other capillary phenomena over large length scales.
\end{abstract}
\section{Introduction}
\href{XX}{\lq\lq Spontaneous Capillarity-Driven Droplet Ejection"} is a short video that provides a demonstration of the auto-ejection of a liquid from a tube under the influence of capillary forces alone. Attention is given to the historical significance and potential research applications of auto-ejection in terrestrial and low-g environments.

NASA scientist R. Siegel was the first to ponder auto-ejection from cylindrical tubes \cite{Siegel1961}. We repeat Siegel's experiments. Footage of the experiment shows the liquid meniscus rise up the partially submerged tube, pin at the lip of the tube, invert, retract, and remain pinned at the tube's end. Siegel concluded that auto-ejection was not possible. De Gennes et al. \cite{DeGennes2004} use a pressure argument to confirm Siegel's results. However by exploiting tube geometry we demonstrate liquids can in fact auto-eject from a tube's end provided sufficient inertia is generated to overpower surface tension forces.

This article provides a brief summary of scaling arguments used to identify critical parametric values for ejection, experiment setup, and results. Scenes from the video are also discussed. Details are provided in \cite{Wollman2012, Wollman2012c}.
\section{Analysis}
\label{sec:DE_Conditions}
\begin{figure*}[ht]
\centering
\def\svgheight{\textwidth}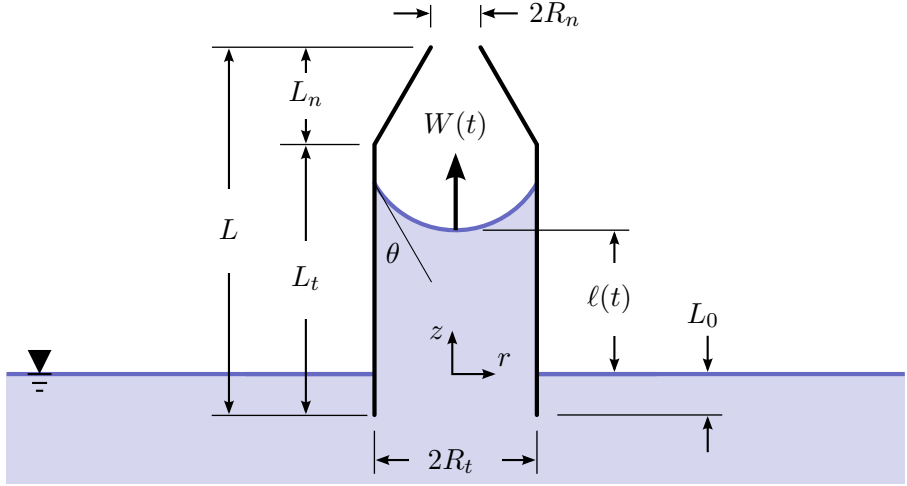
\caption[Schematic of tube with nozzle]{Schematic of tube with nozzle.}
\label{fig:FBD_DE_General}
\end{figure*}

Figure~\ref{fig:FBD_DE_General} introduces the nomenclature of the problem. The fluid wets the interior walls of the partially submerged tube creating a pressure drop. In the absence of gravity the fluid is \lq pumped' along the tube, accelerates in the nozzle and if sufficient velocity is achieved, can eject from the tube end.

\begin{figure*}[ht]
\centering
\def\svgwidth{\textwidth}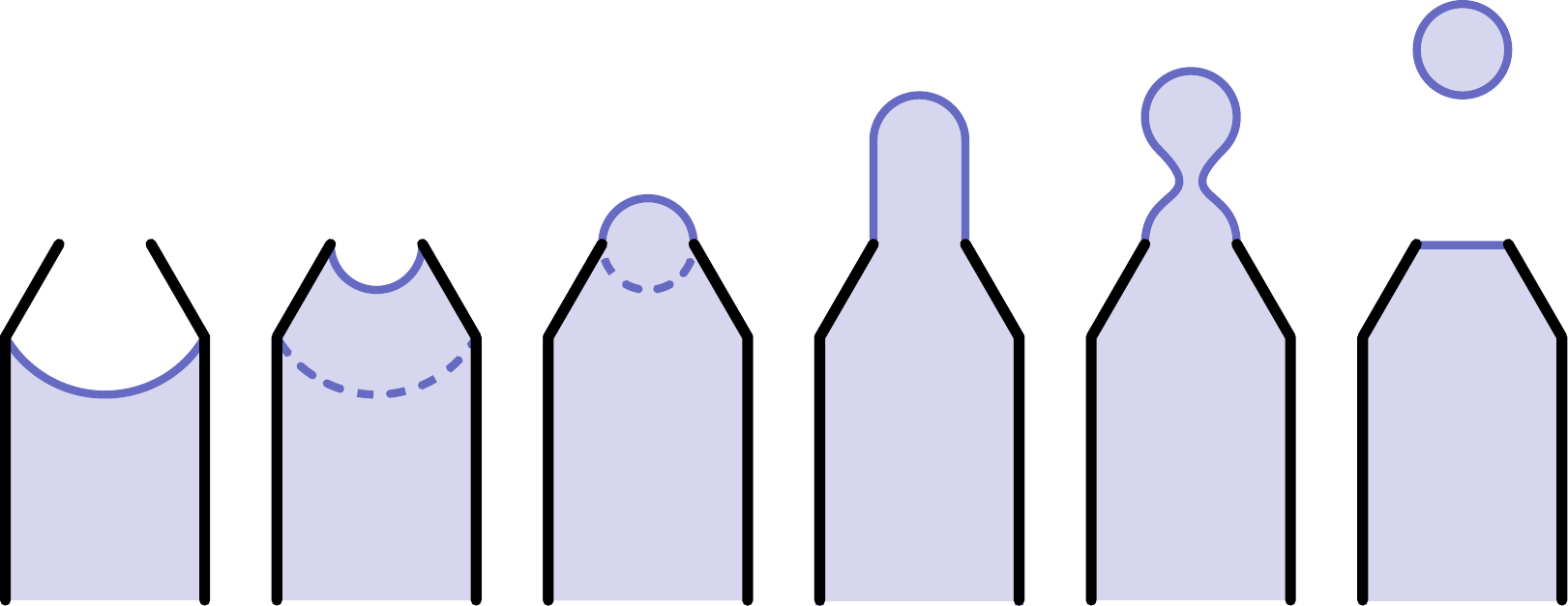
\caption[Schematic of nozzle ejection conditions]{Schematic illustrating the conditions necessary for droplet ejection. (a1) Flow entering the nozzle must (b3) leave inertial, (c) inertia must be able to invert the meniscus from (3) to (4), and (d) remaining inertial forces must overpower surface tension forces sufficient to exceed the (e) Rayleigh breakup length, where (f) at least one drop is pinched off.}
\label{fig:FBD_DE_NozzleDetail}
\end{figure*}

When $t_r/t_\mu \ll 1$, where $t_r\sim\forall_n/Q_t$ is the residence time of liquid in the nozzle, $t_\mu \sim \rho R_{\mathrm{avg}}^2/\mu$ is the viscous diffusion time of the flow, $\forall_n$ is the volume of the nozzle, $Q_t$ is the flow rate entering the nozzle, $R_{\mathrm{avg}}$ is the average nozzle radius, and $\mu$ is the dynamic viscosity, all complexities of the flow in the nozzle due to developing boundary layers, significant viscous normal stresses leading to large dynamic contact angles, and capillary wave dynamics can be ignored and the constricting flow through the nozzle can be assumed to be laminar and inviscid.
Figure~\ref{fig:FBD_DE_NozzleDetail} depicts events of ejection we consider for analysis.

Since the velocity of the meniscus in Figure~\ref{fig:FBD_DE_NozzleDetail}a at position 1 $W_{t1}$ is relatively simple to measure and model \cite{Stange2003}, the critical condition for ejection is written in terms of said velocity.
As the flow accelerates through the nozzle the pressure decreases (See Figure~\ref{fig:FBD_DE_NozzleDetail}a-b). Applying continuity, the flow rate at each end of the nozzle must balance such that
\begin{equation}
W_{n3} = \dfrac{W_{t1}}{\alpha^2\left(1+K_n \right)^{1/2}},
\label{eq:NozzleVelocity2}
\end{equation}
where $K_n$ is the model loss coefficient ascribed to the nozzle and $\alpha = R_n/R_t$.
The meniscus must invert as depicted in Figure~\ref{fig:FBD_DE_NozzleDetail}c. The accompanying increase in pressure results in a reduction in velocity from position 3 to 4 given by
\begin{equation}
 W^2_{n4} = W^2_{n3}- \dfrac{8 \sigma}{\rho R_n},
\label{eq:DymPressLoss2}
\end{equation}
where $\rho$ amd $\sigma$ are the fluid density and surface tension, respectively.
Substituting \ref{eq:NozzleVelocity2} into \ref{eq:DymPressLoss2} yields
\begin{equation}
W_{n4} = \left(\dfrac{W_t1^2}{\alpha^4\left(1+K_n \right)} - \dfrac{8\sigma}{\rho R_n} \right)^{1/2}.
\label{eq:condition2}
\end{equation}
Velocities below this level \lq cannot' invert in the nozzle. After the inversion the flow must still have sufficient inertia to overpower the surface tension force that resists the continued flow required to extend past the Rayleigh breakup length $\sim2\pi R_n$. Balancing inertial and surface tension forces at the end of the nozzle yields the condition above which ejection is expected:
\begin{equation}
\dfrac{\rho R_n W_{n4}^2}{4 \sigma} \gtrsim 1.
\label{eq:condition3}
\end{equation}
Substitution of \ref{eq:condition2} into \ref{eq:condition3} yields
\begin{equation}
\dfrac{\rho R_n}{4 \sigma}\left( \dfrac{W_{t1}^2}{\alpha^4\left(1+K_n \right)} - \dfrac{8\sigma}{\rho R_n} \right)\gtrsim 1.
\label{eq:NotSimplified}
\end{equation}
Introducing modified Weber number $\WE = \rho R_t W_{t1}^2/\sigma\alpha^4\left(1+K_n \right)$, Eq.~\ref{eq:NotSimplified} reduces to
\begin{equation}
\WE\gtrsim12,
\label{eq:Condition3}
\end{equation}
revealing a condition necessary for auto-ejection to occur. Similar scaling of equation~\ref{eq:condition2} reveals that the condition $\WE\lesssim9$ is not likely to eject.
\section{Experiments}
\label{sec:DE_Experimental}
The \href{XX}{video} footage shows a selection of the 200 experiments we conducted using Portland State University's 2.1s Dryden Drop Tower. Additional footage of experiments in a terrestrial lab and aboard the International Space Station(ISS) are also shown. We provide brief descriptions of the experiments' setup and procedures below.
\subsection{Drop tower experiments}
\begin{figure}[ht]
\centering
\subfigure[]{
\def\svgwidth{2in}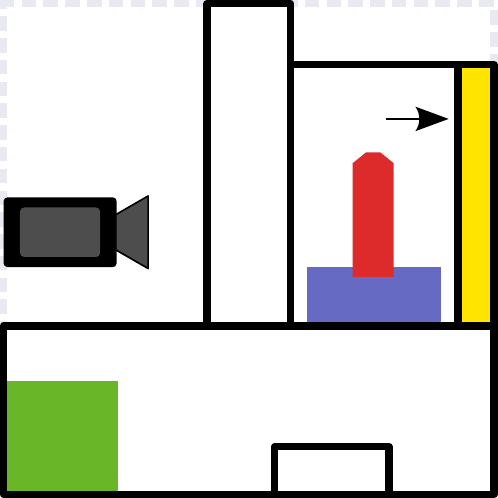
\label{fig:schematic_exp_dt}
}\hspace{10mm}
\subfigure[]{
\includegraphics[height=2in]{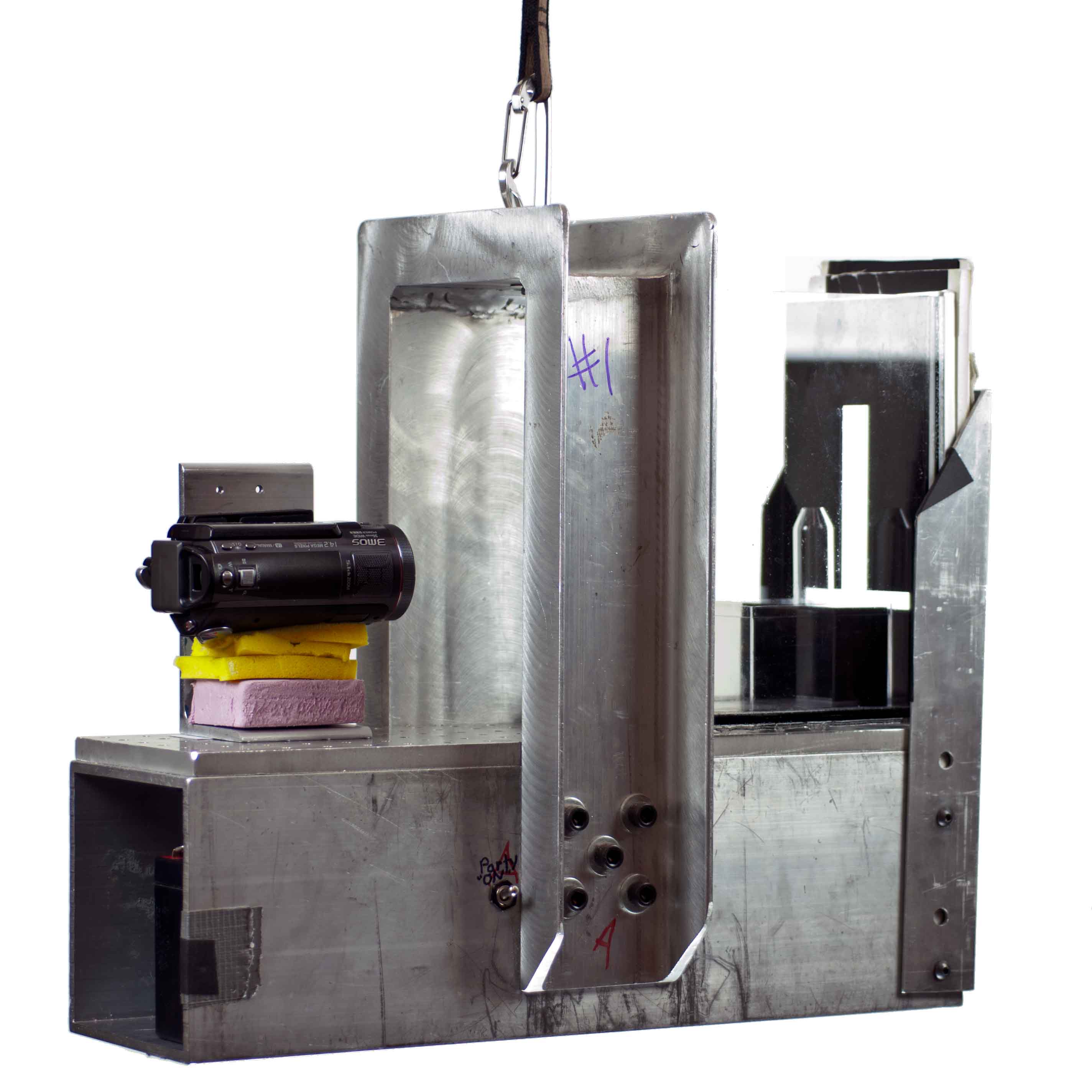}
\label{fig:photo_exp_dt}
}
\caption{Drop tower experiment: (a) schematic and (b) photograph. At one end of the (1) experiment rig main body a (2) camera is mounted. At the opposite end sits the (3) light panel and (4) splash shield. Inside the splash shield sits the (5) fluid reservoir and (6) light guide. The (7) glass tubes are partially submerged in the reservoir. An (8) onboard battery powers the light panel and (9) weights are adjusted to assure a level platform.}
\label{fig:exp_dt}
\end{figure}

Figure \ref{fig:exp_dt} shows an annotated schematic and photograph of the drop tower experiment rig. Simax Glass tubes are employed with nozzles formed by heating, pulling, and grinding. The tubes are partially submerged in a liquid reservoir which is mounted to the experiment rig. The experiment rig is released for 2.1s of freefall inside a drag shield within the drop tower. Cameras image the ensuing capillary rise at 60fps or 400fps. Sample footage is provided in the \href{XX}{video}.
\subsection{$1\g$ Experiments}
\begin{figure}[ht]
\centering
\subfigure[]{
\def\svgwidth{2in}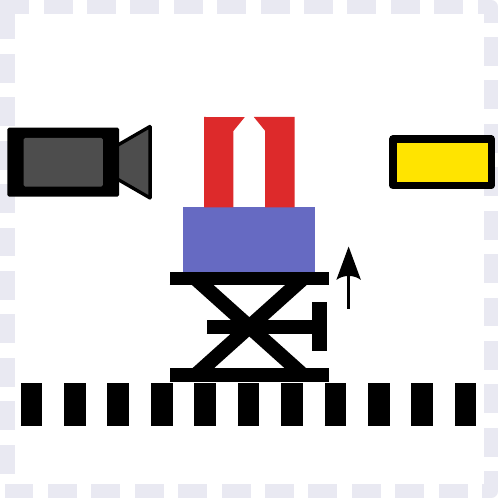
\label{fig:schematic_exp_1g}
}
\subfigure[]{
\includegraphics[height=2in]{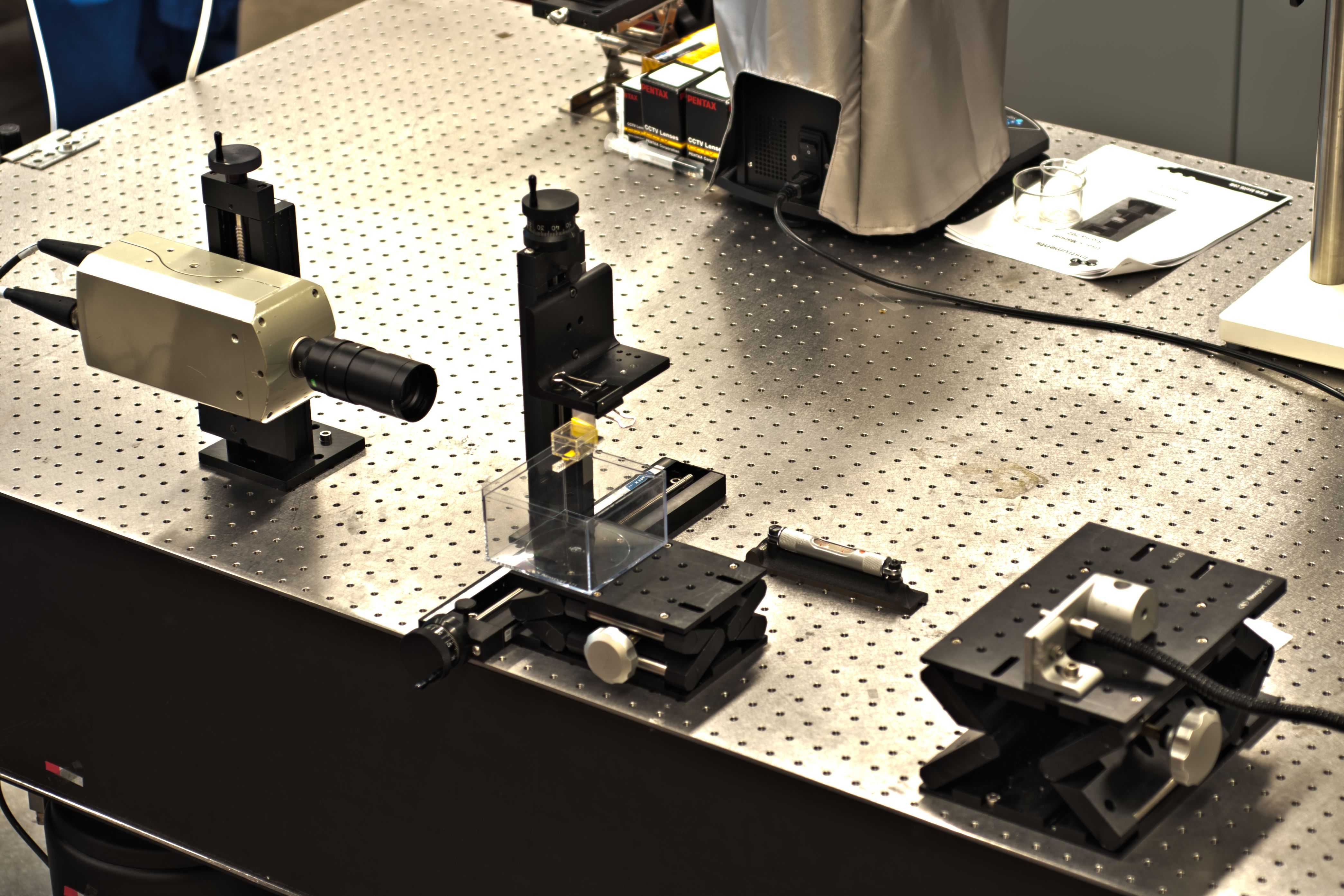}
\label{fig:photo_exp_1g}
}
\caption[]{
$1\g$ experiment setup: (a) schematic and (b) photograph. The experiment is set up on an (4) isolation table using a (1) high speed camera opposite a (5) parallel light source. The (6) acrylic block is mounted between the camera and light source. Below the block the (2) reservoir sits on top of a (3) precision lab jack. }
\label{fig:exp_1g}
\end{figure}

Figure \ref{fig:exp_1g} shows an annotated schematic and photograph of the $1\g$ experiment setup. Hardware for the experiments include a high speed camera, parallel light source, fluid reservoir, precision lab jack, vibration isolation table, and tubes with nozzles machined into acrylic blocks. The machined acrylic block is suspended above the fluid reservoir that sits on top of the lab jack. The jack is slowly raised toward the cylindrical capillary pores. \href{XX}{Video} footage shows the liquid make contact with the block, wick along the base of the block, rise up the pores, accelerate through the nozzles, and auto-eject.
\subsection{ISS experiments}
\begin{figure}[ht]
\centering
\subfigure[]{
\def\svgwidth{2in}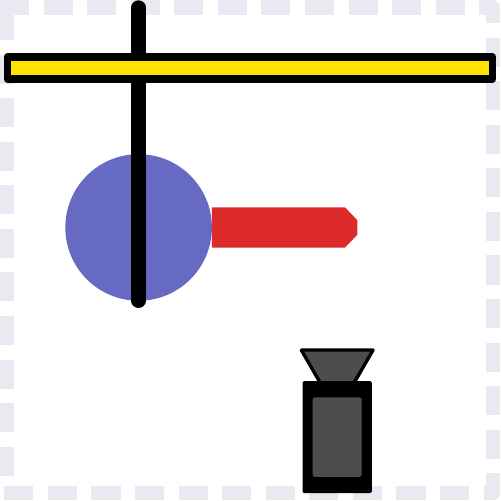
\label{fig:schematic_exp_iss}
}
\subfigure[]{
\includegraphics[height=2in]{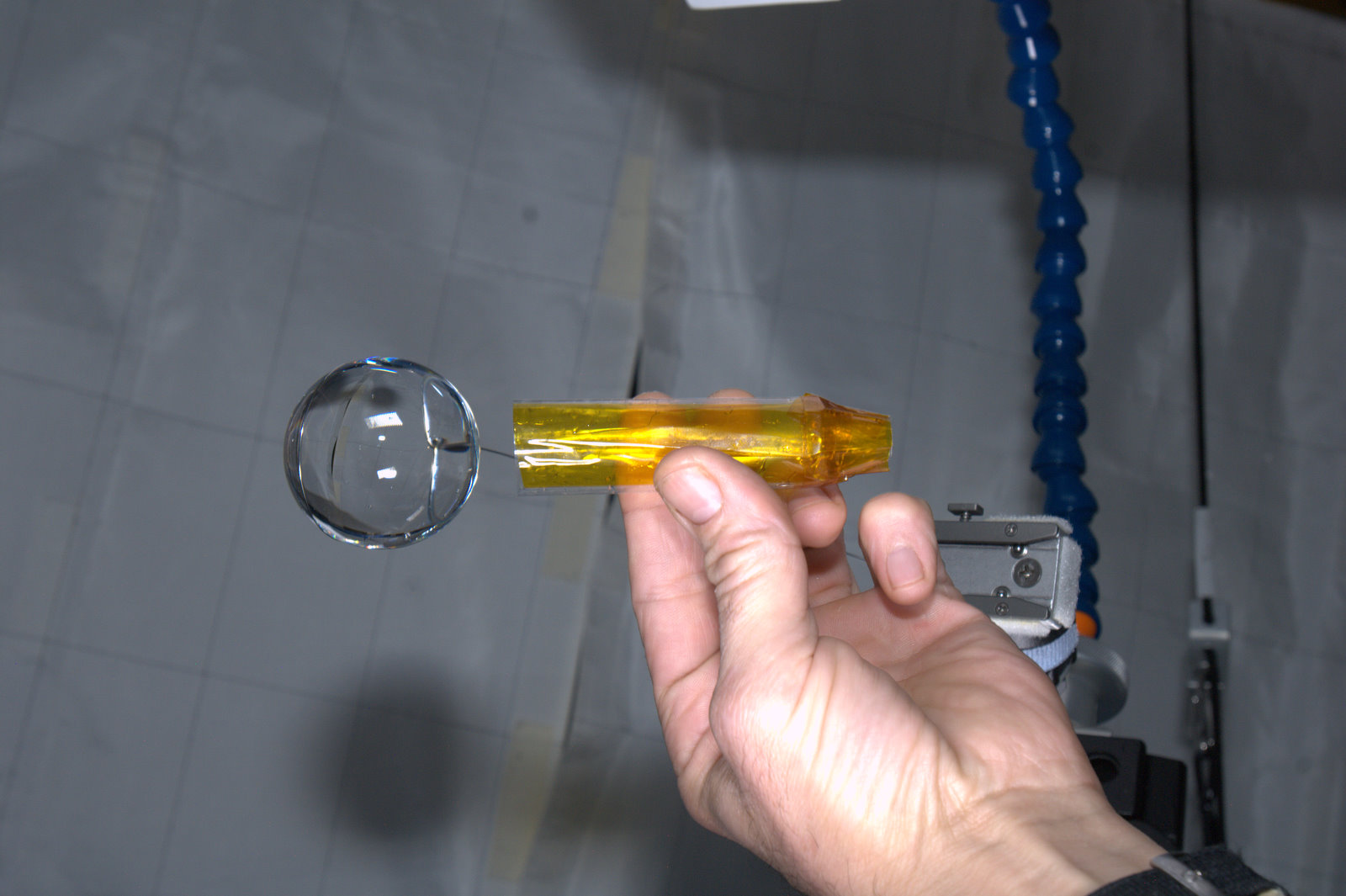}
\label{fig:photo_exp_iss}
}
\caption[]{
Space experiment setup aboard the ISS: (a) schematic and (b) photograph. The (1) camera is mounted orthogonally to (2) a white sheet of paper. The (3) spherical reservoir is actually a large $\sim1.5$l drop and is held in place by (4) a wire loop. The (5) tube is slowly brought into contact with the spherical \lq reservoir'. }
\label{fig:exp_iss}
\end{figure}

Figure \ref{fig:exp_iss} shows an annotated schematic and photograph of the experiment setup used by United States Astronaut Don Pettit who demonstrated auto-ejection of water aboard the ISS. The demonstration includes a polymer sheet tube, a piece of paper, a camera, and a wire loop. The \href{XX}{video} footage shows Pettit slowly bringing the tube into contact with the spherical reservoir and the ensuing capillary \lq rise' and auto-ejection. The water droplet ejected aboard the ISS is $10^6$ times larger than the droplet ejected in the terrestrial experiment.
\section{Results}
\label{sec:DE_Results}
\begin{figure*}
\centering
\def\svgwidth{\textwidth}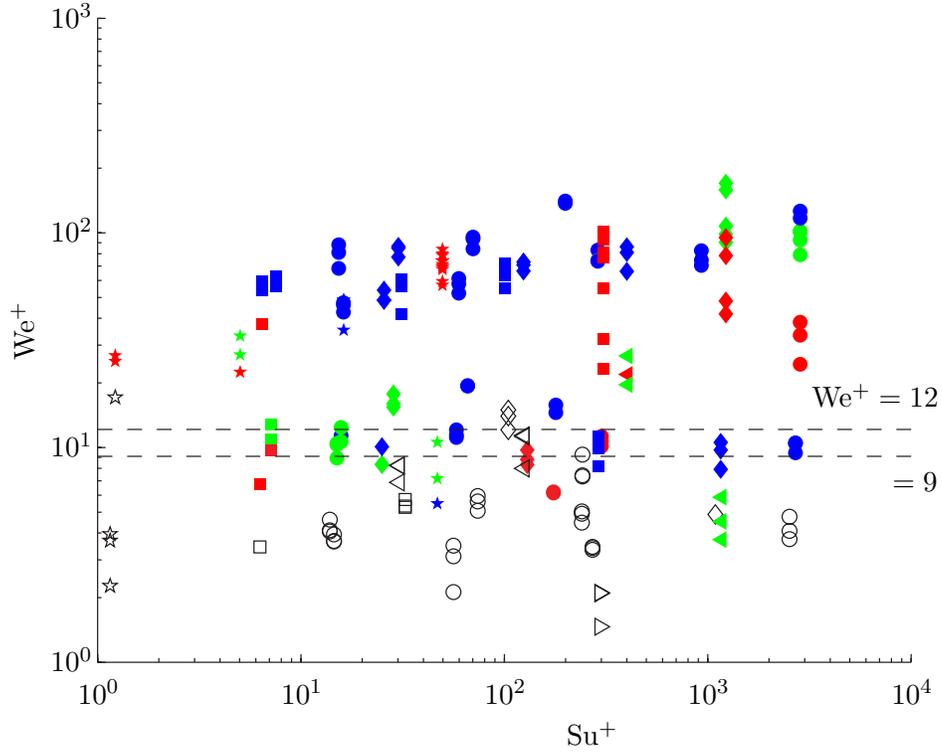
\caption[$\WE$ vs $\SU$]{ $\WE$ vs $\SU$ for droplet auto-ejection. For values of $\WE\gtrsim12$ capillarity-driven droplet ejection is expected. For $\WE\lesssim9$ it is not. See Table \ref{tab:PlotLegend} for symbol meaning.}
\label{fig:Plot_RegimeMap}
\end{figure*}

Figure~\ref{fig:Plot_RegimeMap} is a map for auto-ejection in terms of parameters modified Suratman number $\SU = \rho\sigma R_t^2 / 8 \mu^2 L_t^2 $ and modified Weber number $\WE$. Table~\ref{tab:PlotLegend} is the legend for Figure~\ref{fig:Plot_RegimeMap}. Experiments show fair agreement with the scaling arguments as ejection primarily occurs when $\WE\gtrsim12$ and is essentially absent for $\WE\lesssim9$.
\begin{table}[htbp]
  \centering
  \caption{Plot legend for Figure~\ref{fig:Plot_RegimeMap}}
    \begin{tabular}{cr}
    \toprule
    \multicolumn{1}{c}{Symbol} & \multicolumn{1}{c}{Meaning} \\
    \midrule
    {\color{black} $\circ$} & 0 droplets ejected \\
    {\color{red} $\huge\bullet$} & 1 droplet ejected \\
    {\color{green} $\bullet$} & 2 droplets ejected \\
    {\color{blue} $\bullet$} & $\geq3$ droplets ejected \\
      &  \\
    {\color{blue} $\bullet$} & $\nu = 0.65$cS \\
    {\color{blue} $\blacklozenge$} & $1$cS \\
    {\color{blue} $\blacksquare$} & $ 2$cS \\
    {\color{blue} $\bigstar$} & $5$cS \\
    {\color{blue} $\blacktriangleleft$} & $10$cS \\
    {\color{blue} $\blacktriangleright$} & $20$cS \\
    \bottomrule
    \end{tabular}%
  \label{tab:PlotLegend}%
\end{table}
\section{Additional Footage Explained}
The \href{XX}{video} footage shows a single drop tower experiment illustrating the variety of auto-ejection possibilities. Four $10\mm$ ID tubes partially submerged $L_0=10\mm$ in 0.65cS PDMS with four different nozzles auto-eject jets (which break up into 6, 4, and 2 droplets) and a single droplet.

To illustrate repeatability the \href{XX}{video} shows 10 experiments of nearly identical conditions side by side. A $20.3\mm$ ID tube with a $5\mm$ ID nozzle is partially submerged $L_0=5\mm$ in a 5cS PDMS reservoir. The ejected drop volumes from each experiment are $\forall_{\mathrm{drop}} = 2.11\pm 0.14\ml$.

The \href{XX}{video} concludes with a montage of sample applications of the auto-ejection technique. The first three clips show a cartridge with a 10 x 10 array of 5mm tubes partially submerged in test liquid. The clips show $>300$ auto-ejected drops shoot up into the air, impact a flat smooth surface (rebounds abound), and impact a textured surface designed to capture and hold the droplets. Work is continuing in this direction.

To illustrate how auto-ejection is viable for droplet combustion experiments, a fourth clip shows an asymmetric u-tube with a nozzle pointed horizontally at a candle. The flammable 0.65cS PDMS liquid rises in the smaller tube with the higher curvature pushing a column of fuel vapor ahead of it. The candle back-ignites the combustible vapor and the flame resides at the nozzle exit. The spontaneously ejected droplets that follow ignite and fly through the air leaving a trail of soot behind them. One drop is small enough to self-extinguish.

\section{Conclusion}
\label{sec:DE_Conclusion}
Contrary to conventional wisdom, capillarity-driven droplet ejection is possible, predictable, repeatable, and a viable method for drop-on-demand delivery requirements for further capillary fluidic research. The \href{XX}{video} shows only a few of the many examples of potential applications for auto-ejection. The technique described herein can be used as a means to study drop formation, drop to jet transition, jet breakup, and droplet combustion at larger characteristic lengths then are achievable at 1\g. Droplet impact, splash, rebound, satellites, adhesion and coalescence are ripe fields that could benefit from this approach.
\bibliographystyle{unsrt}
\addcontentsline{toc}{section}{\numberline{}References}
\bibliography{SpontaneousCapillarityDrivenDropletEjection_APS_GFM_Wollman}
\end{document}